\documentclass[12pt]{article}
\usepackage{epsfig}
\usepackage{amsmath}
\usepackage{hhline}
\usepackage{amssymb}
\usepackage{times}
\usepackage{cite}
\usepackage{colordvi}

\newlength{\dinwidth}
\newlength{\dinmargin}
\begin{document}  
\newcommand{\pom}{{I\!\!P}}
\newcommand{\reg}{{I\!\!R}}
\newcommand{\slowpi}{\pi_{\mathit{slow}}}
\newcommand{\fiidiii}{F_2^{D(3)}}
\newcommand{\fiidiiiarg}{\fiidiii\,(\beta,\,Q^2,\,x)}
\newcommand{\n}{1.19\pm 0.06 (stat.) \pm0.07 (syst.)}
\newcommand{\nz}{1.30\pm 0.08 (stat.)^{+0.08}_{-0.14} (syst.)}
\newcommand{\fiidiiiful}{F_2^{D(4)}\,(\beta,\,Q^2,\,x,\,t)}
\newcommand{\fiipom}{\tilde F_2^D}
\newcommand{\ALPHA}{1.10\pm0.03 (stat.) \pm0.04 (syst.)}
\newcommand{\ALPHAZ}{1.15\pm0.04 (stat.)^{+0.04}_{-0.07} (syst.)}
\newcommand{\fiipomarg}{\fiipom\,(\beta,\,Q^2)}
\newcommand{\pomflux}{f_{\pom / p}}
\newcommand{\nxpom}{1.19\pm 0.06 (stat.) \pm0.07 (syst.)}
\newcommand {\gapprox}
   {\raisebox{-0.7ex}{$\stackrel {\textstyle>}{\sim}$}}
\newcommand {\lapprox}
   {\raisebox{-0.7ex}{$\stackrel {\textstyle<}{\sim}$}}
\def\gsim{\,\lower.25ex\hbox{$\scriptstyle\sim$}\kern-1.30ex%
\raise 0.55ex\hbox{$\scriptstyle >$}\,}
\def\lsim{\,\lower.25ex\hbox{$\scriptstyle\sim$}\kern-1.30ex%
\raise 0.55ex\hbox{$\scriptstyle <$}\,}
\newcommand{\pomfluxarg}{f_{\pom / p}\,(x_\pom)}
\newcommand{\dsf}{\mbox{$F_2^{D(3)}$}}
\newcommand{\dsfva}{\mbox{$F_2^{D(3)}(\beta,Q^2,x_{I\!\!P})$}}
\newcommand{\dsfvb}{\mbox{$F_2^{D(3)}(\beta,Q^2,x)$}}
\newcommand{\dsfpom}{$F_2^{I\!\!P}$}
\newcommand{\gap}{\stackrel{>}{\sim}}
\newcommand{\lap}{\stackrel{<}{\sim}}
\newcommand{\fem}{$F_2^{em}$}
\newcommand{\tsnmp}{$\tilde{\sigma}_{NC}(e^{\mp})$}
\newcommand{\tsnm}{$\tilde{\sigma}_{NC}(e^-)$}
\newcommand{\tsnp}{$\tilde{\sigma}_{NC}(e^+)$}
\newcommand{\st}{$\star$}
\newcommand{\sst}{$\star \star$}
\newcommand{\ssst}{$\star \star \star$}
\newcommand{\sssst}{$\star \star \star \star$}
\newcommand{\tw}{\theta_W}
\newcommand{\sw}{\sin{\theta_W}}
\newcommand{\cw}{\cos{\theta_W}}
\newcommand{\sww}{\sin^2{\theta_W}}
\newcommand{\cww}{\cos^2{\theta_W}}
\newcommand{\trm}{m_{\perp}}
\newcommand{\trp}{p_{\perp}}
\newcommand{\trmm}{m_{\perp}^2}
\newcommand{\trpp}{p_{\perp}^2}
\newcommand{\alp}{\alpha_s}

\newcommand{\alps}{\alpha_s}
\newcommand{\sqrts}{$\sqrt{s}$}
\newcommand{\LO}{$O(\alpha_s^0)$}
\newcommand{\Oa}{$O(\alpha_s)$}
\newcommand{\Oaa}{$O(\alpha_s^2)$}
\newcommand{\PT}{p_{\perp}}
\newcommand{\JPSI}{J/\psi}
\newcommand{\sh}{\hat{s}}
\newcommand{\uh}{\hat{u}}
\newcommand{\MP}{m_{J/\psi}}
\newcommand{\PO}{I\!\!P}
\newcommand{\xbj}{x}
\newcommand{\xpom}{x_{\PO}}
\newcommand{\ttbs}{\char'134}
\newcommand{\xpomlo}{3\times10^{-4}}  
\newcommand{\xpomup}{0.05}  
\newcommand{\dgr}{^\circ}
\newcommand{\pbarnt}{\,\mbox{{\rm pb$^{-1}$}}}
\newcommand{\gev}{\,\mbox{GeV}}
\newcommand{\WBoson}{\mbox{$W$}}
\newcommand{\fbarn}{\,\mbox{{\rm fb}}}
\newcommand{\fbarnt}{\,\mbox{{\rm fb$^{-1}$}}}
\newcommand{\dsdx}[1]{$d\sigma\!/\!d #1\,$}
\newcommand{\eV}{\mbox{e\hspace{-0.08em}V}}
%
%
\newcommand{\qsq}{\ensuremath{Q^2} }
\newcommand{\gevsq}{\ensuremath{\mathrm{GeV}^2} }
\newcommand{\et}{\ensuremath{E_t^*} }
\newcommand{\rap}{\ensuremath{\eta^*} }
\newcommand{\gp}{\ensuremath{\gamma^*}p }
\newcommand{\dsiget}{\ensuremath{{\rm d}\sigma_{ep}/{\rm d}E_t^*} }
\newcommand{\dsigrap}{\ensuremath{{\rm d}\sigma_{ep}/{\rm d}\eta^*} }

\newcommand{\dstar}{\ensuremath{D^*}}
\newcommand{\dstarp}{\ensuremath{D^{*+}}}
\newcommand{\dstarm}{\ensuremath{D^{*-}}}
\newcommand{\dstarpm}{\ensuremath{D^{*\pm}}}
\newcommand{\zDs}{\ensuremath{z(\dstar )}}
\newcommand{\Wgp}{\ensuremath{W_{\gamma p}}}
\newcommand{\ptds}{\ensuremath{p_t(\dstar )}}
\newcommand{\etads}{\ensuremath{\eta(\dstar )}}
\newcommand{\ptj}{\ensuremath{p_t(\mbox{jet})}}
\newcommand{\ptjn}[1]{\ensuremath{p_t(\mbox{jet$_{#1}$})}}
\newcommand{\etaj}{\ensuremath{\eta(\mbox{jet})}}
\newcommand{\detadsj}{\ensuremath{\eta(\dstar )\, \mbox{-}\, \etaj}}

\def\Journal#1#2#3#4{{#1} {\bf #2} (#3) #4}
\def\NCA{\em Nuovo Cimento}
\def\NIM{\em Nucl. Instrum. Methods}
\def\NIMA{{\em Nucl. Instrum. Methods} {\bf A}}
\def\NPB{{\em Nucl. Phys.}   {\bf B}}
\def\PLB{{\em Phys. Lett.}   {\bf B}}
\def\PRL{\em Phys. Rev. Lett.}
\def\PRD{{\em Phys. Rev.}    {\bf D}}
\def\ZPC{{\em Z. Phys.}      {\bf C}}
\def\EJC{{\em Eur. Phys. J.} {\bf C}}
\def\CPC{\em Comp. Phys. Commun.}

\newcommand{\be}{\begin{equation}} 
\newcommand{\ee}{\end{equation}} 
\newcommand{\ba}{\begin{eqnarray}} 
\newcommand{\ea}{\end{eqnarray}}

\begin{titlepage}

\noindent
\begin{flushleft}
{\tt DESY 09-108    \hfill    ISSN 0418-9833} \\
{\tt July 2009}                  \\
\end{flushleft}

\noindent

\vspace{2cm}
\begin{center}
\begin{Large}

{\bf Multi-Leptons with High Transverse Momentum at HERA \\}

\vspace{2cm}

The H1 and ZEUS Collaborations

\end{Large}
\end{center}

\vspace{2cm}

\begin{abstract}

Events with at least two high transverse momentum leptons (electrons or muons) are studied using the H1 and ZEUS detectors at HERA with an integrated luminosity of $0.94$~fb$^{-1}$.
The observed numbers of events are in general agreement with the Standard Model predictions. 
Seven di- and tri-lepton events are observed in $e^+p$ collision data with a scalar sum of the lepton transverse momenta above $100$~GeV  while \mbox{$1.94 \pm 0.17$} events are expected. 
Such events are not observed in $e^-p$ collisions  for which $1.19 \pm 0.12$ are predicted.
Total visible and differential di-electron and di-muon photoproduction cross sections are extracted in a restricted phase space dominated by photon-photon collisions.

\end{abstract}

\vspace{1.5cm}

\begin{center}
Accepted by JHEP
\end{center}

\end{titlepage}

\begin{flushleft}
  \begin{center}                                                                                     
{                      \Large  The H1 and ZEUS Collaborations              }                               
\end{center}                                                                                       
F.D.~Aaron$^{13, a8}$,
H.~Abramowicz$^{72, a37}$,
I.~Abt$^{57}$,
L.~Adamczyk$^{19}$,
M.~Adamus$^{84}$,
M.~Aldaya~Martin$^{31}$,
C.~Alexa$^{13}$,
K.~Alimujiang$^{31}$,
V.~Andreev$^{54}$,
S.~Antonelli$^{9}$
% ...

{\it
\vspace{0.4cm}

 $^{1}$  I. Physikalisches Institut der RWTH, Aachen, Germany~$^{b1}$ \\
% ...
$^{9}$   {\it University and INFN Bologna, Bologna, Italy}~$^{b17}$ \\
% ...
$^{13}$  National Institute for Physics and Nuclear Engineering (NIPNE), Bucharest, Romania \\
% ...
$^{17}$   {\it Calabria University, Physics Department and INFN, Cosenza, Italy}~$^{b17}$ \\
% ...
$^{19}$   {\it Faculty of Physics and Applied Computer Science, AGH-University of Science and \mbox{Technology}, Cracow, Poland}~$^{b27}$ \\
% ...
$^{31}$   {\it Deutsches Elektronen-Synchrotron DESY, Hamburg, Germany} \\
% ...
$^{54}$  Lebedev Physical Institute, Moscow, Russia~$^{b6}$ \\
% ...
$^{57}$   {\it Max-Planck-Institut f\"ur Physik, M\"unchen, Germany} \\
% ...
$^{72}$   {\it Raymond and Beverly Sackler Faculty of Exact Sciences, School of Physics, Tel Aviv University, Tel Aviv, Israel}~$^{b16}$ \\
% ...
$^{84}$   {\it Institute for Nuclear Studies, Warsaw, Poland} \\
 $^{85}$  Fachbereich C, Universit\"at Wuppertal, Wuppertal, Germany \\
 $^{86}$  Yerevan Physics Institute, Yerevan, Armenia \\
 $^{87}$   {\it Meiji Gakuin University, Faculty of General Education, Yokohama, Japan}~$^{b18}$ \\
 $^{88}$   {\it Department of Physics, York University, Ontario, Canada M3J1P3}~$^{b14}$ \\
 $^{89}$   {\it Deutsches Elektronen-Synchrotron DESY, Zeuthen, Germany} \\
 $^{90}$  Institut f\"ur Teilchenphysik, ETH, Z\"urich, Switzerland~$^{b10}$ \\
 $^{91}$  Physik-Institut der Universit\"at Z\"urich, Z\"urich, Switzerland~$^{b10}$ \\

\vspace{0.4cm}
{ \small


$^{a1}$  Also at Physics Department, National Technical University, Zografou Campus, GR-15773 Athens, Greece \\
$^{a2}$  Also at Rechenzentrum, Universit\"at Wuppertal, Wuppertal, Germany \\
$^{a3}$  Also at University of P.J. \v{S}af\'arik, Ko\v{s}ice, Slovak Republic \\
$^{a4}$  Also at CERN, Geneva, Switzerland \\
$^{a5}$  Also at Max-Planck-Institut f\"ur Physik, M\"unchen, Germany \\
$^{a6}$  Also at Comenius University, Bratislava, Slovak Republic \\
$^{a7}$  Also at DESY and University Hamburg, Helmholtz Humboldt Research Award \\
$^{a8}$  Also at Faculty of Physics, University of Bucharest, Bucharest, Romania \\
% ...
$^{a37}$   Also at Max Planck Institute, Munich, Germany, Alexander von Humboldt Research Award\\
% ...
$^{b1}$  Supported by the German Federal Ministry for Education and Research (BMBF), under contract numbers 05 H1 1GUA /1, 05 H1 1PAA /1, 05 H1 1PAB /9, 05 H1 1PEA /6, 05 H1 1VHA /7 and 05 H1 1VHB /5 \\

% ...
$^{b6}$  Supported by the Deutsche Forschungsgemeinschaft \\
% ...
$^{b16}$   Supported by the Israel Science Foundation\\
$^{b17}$   Supported by the Italian National Institute for Nuclear Physics (INFN) \\
% ...
$^{b27}$   Supported by the Polish Ministry of Science and Higher Education as a scientific project (2009-2010)\\
% ...

}
}

\end{flushleft}

\newpage

\section{Introduction}

According to predictions of the Standard Model (SM) the production of multi-lepton final states in electron\footnote{Here and in the following, the term ``electron'' denotes generically both the electron and the positron.}-proton collisions proceeds mainly via photon-photon interactions~\cite{Vermaseren:1982cz}.
The clean experimental signature of leptons with high transverse momenta, $P_T$, together with the precisely calculable small SM cross section provides high sensitivity to possible contributions of physics beyond the SM.
Measurements of multi-lepton production at the HERA collider have already been performed by the H1~\cite{Aktas:2006nu,Aktas:2003jg,Aktas:2003sz,mlep_H1} and ZEUS~\cite{mlep_ZEUS} collaborations using data samples corresponding to an integrated luminosity of  $\sim 0.5$~fb$^{-1}$ per experiment.
Events with high invariant mass $M_{12}$ of the two highest $P_T$ leptons or high scalar sum of transverse momenta of all leptons $\sum P_T$ were measured by both experiments in a region where the SM expectation is low.
The yields of multi-lepton events were found to be in general agreement with the SM predictions in both H1 and ZEUS analyses.

A combination of the H1 and ZEUS results which exploits the complete $e^\pm p$ data samples of both experiments is presented in this paper.
Total yields and kinematic distributions of multi-lepton final states with electrons or muons are measured and compared to the SM.
The two-fold increase in the available data statistics allows a more 
stringent test of the SM in the high mass and high $\sum P_T$ regions.
In addition, total visible and differential photoproduction cross sections of $e^+e^-$ and $\mu^+\mu^-$ pairs are measured in a restricted phase-space region dominated by photon-photon collisions. 

The analysed data were collected between $1994$ and $2007$ at the HERA electron-proton collider using the H1 and ZEUS detectors.
The electron and proton beam energies were respectively $27.6$~GeV and $820$~GeV or $920$~GeV, corresponding to centre-of-mass energies $\sqrt{s}$ of $301$~GeV or $319$~GeV.
The data correspond to an integrated luminosity of $0.94$~fb$^{-1}$, comprising  $0.38$~fb$^{-1}$ of $e^-p$ collisions and  $0.56$~fb$^{-1}$ of $e^+p$ collisions, with $8\%$ of the total collected at $\sqrt{s} = 301$~GeV.

The H1 and ZEUS detectors are general purpose instruments which consist of tracking systems surrounded by electromagnetic and hadronic calorimeters and muon detectors, ensuring close to 4$\pi$ coverage of the $ep$ interaction point.
The origin of the coordinate system is the nominal $ep$ interaction point, with the direction of the proton beam defining the positive $z$-axis (forward region). The  $x-y$ plane is called the transverse plane and $\phi$ is the azimuthal angle. The pseudorapidity $\eta$ is defined as $\eta = -\ln \, \tan (\theta/2)$, where $\theta$ is the polar angle. 
Detailed descriptions of the H1 and ZEUS detectors can be found elsewhere~\cite{H1_det,ZEUS_det}.

\section{Experimental Method}
\label{sec:selection}

For this analysis, a common phase-space region is chosen according to the individual performances of the H1 and ZEUS detectors, such that both detectors have high and well understood acceptance.
The common phase-space region is somewhat smaller than those used by the respective collaborations~\cite{mlep_H1,mlep_ZEUS} and is described in the following.

The event selection proceeds in two steps. Electron or muon candidates are first identified using a wider angular range and lower energy thresholds allowed by the detectors. 
In a second step, in order to minimise the background present in some of the event topologies, at least two central ($20^\circ < \theta < 150^\circ$) lepton candidates are required.

Electron candidates are identified in the polar-angle range $5^\circ < \theta < 175^\circ$ as compact and isolated energy deposits in the electromagnetic calorimeters.
The electron energy threshold is $10$~GeV in the range $5^\circ < \theta < 150^\circ$ and $5$~GeV in the backward region $150^\circ < \theta < 175^\circ$.
Compared to the published H1 analysis~\cite{mlep_H1}, the electron energy threshold is here raised in the central region $20^\circ < \theta < 150^\circ$ from $5$ to $10$~GeV.
Muon candidates are identified in the range $20^\circ <\theta < 160^\circ$ with a minimum transverse momentum of $2$~GeV.
Muon identification is based on the measurement of a track in the inner tracking system associated to a track segment reconstructed in the muon chambers or an energy deposit in the calorimeter compatible with a minimum ionising particle.
Only tracks associated with the primary event vertex are used in the analysis.
Detailed descriptions of electron and muon identification criteria used by the H1 and ZEUS experiments are given in the individual publications~\cite{mlep_H1, mlep_ZEUS}.
For the H1 experiment, the resulting electron identification efficiency is $80$\% in the central region and larger than $95$\% in the forward and backward regions, while for the ZEUS detector the electron identification efficiency is  $90$\%.
The lower electron identification efficiency in the H1 analysis is mainly due to a tight matching requirement between the transverse momenta measured by the tracker and the calorimeter~\cite{Aktas:2003jg,mlep_H1}.
The efficiency to identify muons in the H1 and ZEUS analyses is $90$\% and $55$\%, respectively. 
The lower muon identification efficiency for ZEUS is due to a lower
performance and a smaller fiducial volume of the muon system and
a low efficiency of the track trigger for low multiplicity events~\cite{mlep_ZEUS}.

Multi-lepton events are selected by requiring at least two
central lepton candidates, of which one must have $P_T^\ell > 10$~GeV and the other
$P_T^{\ell} > 5$~GeV. 
Additional leptons identified in the
detector according to the criteria defined above may be present in the event.
All lepton candidates are required to be isolated with respect to each other by a minimum distance of at least $0.5$ units in the $\eta-\phi$ plane.
No explicit requirement on the charge of the lepton candidates is imposed.
Lepton candidates are
ordered according to decreasing transverse momentum, $P_T^{\ell_i} > P_T^{\ell_{i+1}}$.
According to the number and the flavour of the lepton candidates, the events are classified into mutually exclusive topologies. 

The production cross section of $e^+e^-$ and $\mu^+\mu^-$ pairs is measured 
in the photoproduction regime, in which the virtuality $Q^2$ of the photon emitted by the beam electron is low. Subsamples of $ee$ and $\mu\mu$ events, dominated by photon-photon collisions, labelled $(\gamma\gamma)_{e}$ and $(\gamma\gamma)_{\mu}$, are selected by requiring the difference $E -P_z$ between the energy and the longitudinal momentum of all visible particles to be lower than 45~GeV.
This requirement selects events in which the scattered electron is lost in the beampipe and corresponds to cuts on  $Q^2 < 1$~GeV$^2$ and on the event inelasticity, $y= (E -P_z)/2E_e < 0.82$, where $E_e$ is the electron beam energy. 

The GRAPE~\cite{Abe:2000cv} Monte Carlo (MC) event generator is used to calculate SM production cross sections, dominated by photon-photon interactions, $\gamma \gamma \rightarrow \ell^+ \ell^-$, and to simulate multi-lepton events.
GRAPE predicts cross sections for
$ep \rightarrow e \: \mu^+ \mu^- X$ and $ep \rightarrow e \: e^+ e^- X$  processes, leading to $e\mu\mu$ and $eee$ final states.
 Events with only two leptons ($\mu\mu$, $e\mu$ or $ee$) are observed if the scattered electron or one lepton of the pair is not detected. 
The $ep \rightarrow e \: \tau^+ \tau^- X$ process with subsequent leptonic tau decays is also simulated with GRAPE.

Experimental background contributions from various SM processes to the selected multi-lepton topologies were studied~\cite{mlep_H1,mlep_ZEUS}.
Backgrounds to the $ee$  final state arise from neutral current (NC) deep inelastic scattering (DIS) events ($ep \rightarrow e X$) in which, in addition to the scattered electron, hadrons or radiated photons are wrongly identified as electrons, and from QED Compton  (QEDC) events ($ep \rightarrow e \gamma X$) if the photon is misidentified as an electron. 
Background to the $e\mu$ final state arises from NC DIS events if hadrons are misidentified as muons.
The background contributions to $eee$, $e\mu\mu$ and $\mu\mu$ final states are negligible.

The combination of the results of the H1 and ZEUS experiments is performed both on the number of observed events and at the cross section level.
Distributions of data events and of MC expectations are added bin by bin.
Experimental systematic uncertainties are treated as uncorrelated between the experiments.
A detailed list of all experimental systematic uncertainties of both experiments can be found in the individual publications~\cite{mlep_H1, mlep_ZEUS}.
The theoretical uncertainty of $3$\% on the total lepton pair contribution calculated from the GRAPE MC is considered to be correlated between the experiments.
Cross sections measured by H1 and ZEUS are combined using a weighted average~\cite{pdg}.

\section{Results} 

The total number of selected events in the data are compared to SM predictions in Table~\ref{tab:mlepyields} for the $ee$, $\mu\mu$, $e\mu$, $eee$ and $e\mu\mu$ topologies and for the $\gamma\gamma$ subsamples. 
The observed numbers of events are in good agreement with the SM expectations.
The $e\mu\mu$, $\mu\mu$ and $e\mu$ topologies are dominated by muon pair production while the $eee$ and $ee$ topologies contain mainly events from electron pair production.
The contribution from tau pair production is $\sim 4\%$ in the $e\mu$ topology,
negligible in the others, and is considered as signal.
The NC DIS and QEDC processes give rise to a sizeable background contribution in the $ee$ topology where the H1 and ZEUS analyses have slightly different background rejection capabilities. 
The contribution from NC DIS and QEDC processes to the total SM expectation amounts to $24\%$ for ZEUS and $11$\% for H1 due to the tighter electron identification criteria.
Most of the events in the $e\mu$ topology arise from muon pair production at high $Q^2$, in which the beam electron is scattered at a large angle in the detector, while one of the muons is outside the acceptance region.
In this topology, the NC DIS background contributes $\sim 10$\% in both the H1 and ZEUS experiments.
Four events with four lepton candidates  in the final state are also 
observed, 2 $eeee$ and 2 $ee\mu\mu$. 
The total SM expectation for four-lepton events 
is $2.3^{+0.7}_{-0.2}$, including background where a fourth lepton 
is a misidentified hadron. The contribution from true
four-lepton events, originating from higher-order QED processes,
 is not included in the SM and is expected to be
negligible.

\begin{figure}[!htbp] 
  \begin{center}
  \includegraphics[width=\textwidth]{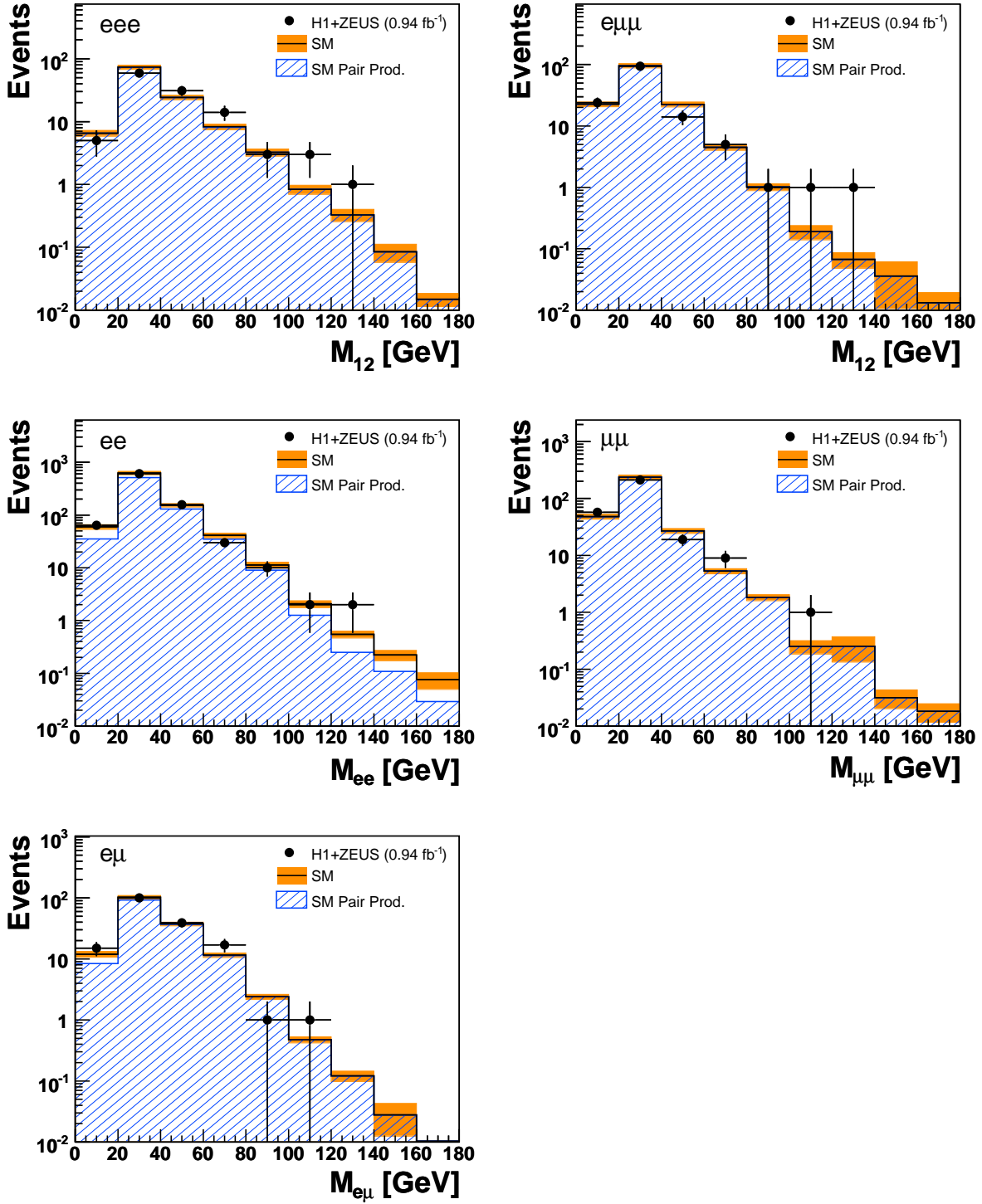}
  \end{center}
  \caption{The distribution of the invariant mass of the two highest $P_T$ leptons for events classified as $eee$, $e\mu\mu$, $ee$, $\mu\mu$ and $e\mu$. 
  The points correspond to the observed data events and the histogram to the SM expectation. The total uncertainty on the SM expectation is given by the shaded band. The component of the SM expectation arising from lepton pair production is given by the hatched histogram.
   }
\label{fig:Masses}  
\end{figure}

The distributions of the invariant mass $M_{12}$ of the two highest $P_T$ leptons for the different topologies are shown in Fig.~\ref{fig:Masses}.
An overall agreement with the SM prediction is observed in all cases.
Events with high invariant mass ($M_{12} > 100$~GeV) are observed in the data. 
The corresponding observed and predicted event yields are summarised for all topologies in Table~\ref{tab:mlepyieldsM100}.
One $ee$ and two $eee$ high mass events are observed by ZEUS~\cite{mlep_ZEUS}. Nine high mass events are observed by H1.
Compared to the H1 results~\cite{mlep_H1}, one $eee$ high mass event is not selected in this combined analysis due to the increased electron energy threshold of $10$~GeV in the central region.
The results for $e^+p$ and $e^-p$ data are also shown separately in Table~\ref{tab:mlepyieldsM100}. 
All high mass events observed by both experiments originate from  $e^+p$ collisions. Several of these events also have high $\sum P_T$ values.

\begin{figure}[htbp] 
\begin{center}
\includegraphics[width=.5\textwidth]{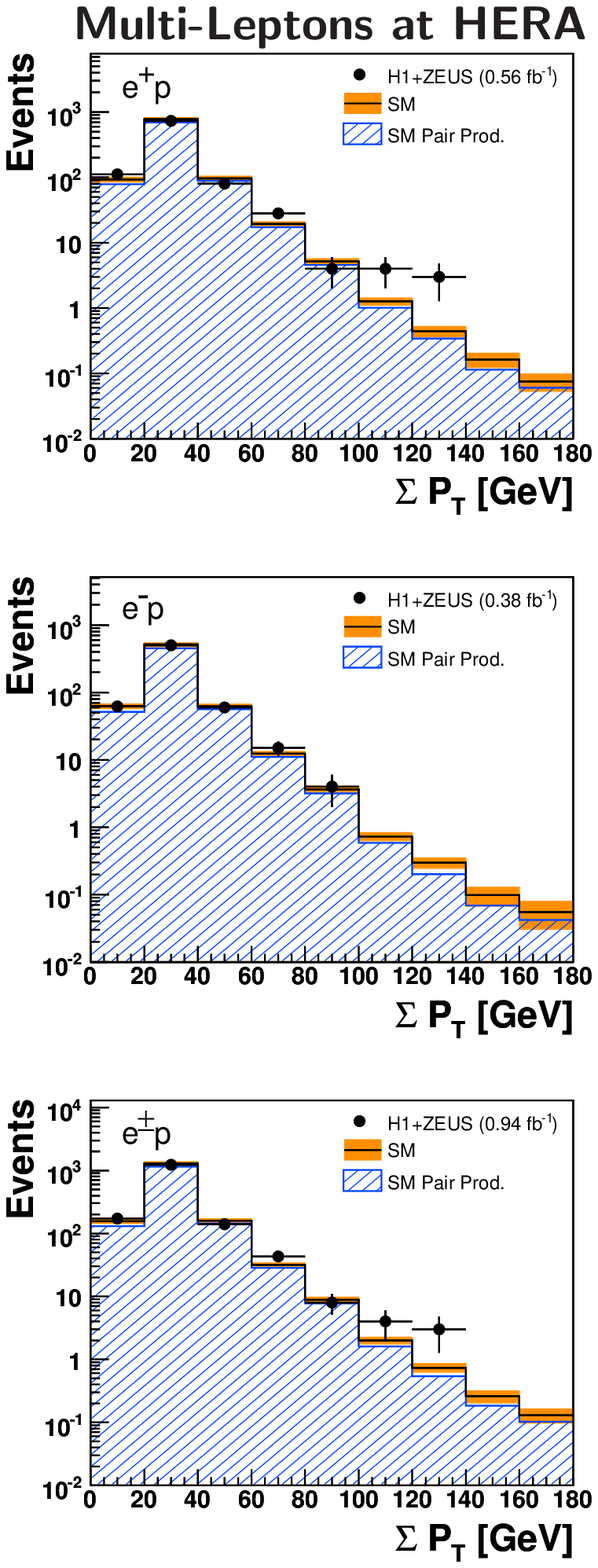}
\end{center}
\vspace{-0.5cm}
  \caption{The distribution of the scalar sum of the transverse momenta $\sum P_T$ for combined di-lepton and tri-lepton event topologies for all data as well as for $e^+p$ and $e^-p$.
  The points correspond to the observed data events and the histogram to the SM expectation. The total uncertainty on the SM expectation is given by the shaded band. The component of the SM expectation arising from lepton pair production is given by the hatched histogram.
}
\label{fig:SumEt_All_lep}
\end{figure}

Figure~\ref{fig:SumEt_All_lep} presents the distributions of $\sum P_T$ of the observed multi-lepton events compared to the SM expectation. 
Good overall agreement between the data and the SM prediction is observed.
For $\sum P_T >$~$100$~GeV, seven events are
observed in total, compared to \mbox{$3.13 \pm 0.26$} expected from the SM (see Table~\ref{tab:mlepyieldsEt100}). 
These seven events were all recorded in the $e^+p$ data, for which the SM expectation is $1.94 \pm 0.17$.
The events correspond to the four $ee$ and the two $e\mu\mu$ events observed with $M_{12} > 100$~GeV, together with one $eee$ event observed with \mbox{$M_{12} = 93$~GeV}.

Total visible and differential cross sections for di-electron and di-muon production are measured using the selected $(\gamma\gamma)_{e}$ and $(\gamma\gamma)_{\mu}$ subsamples.
The kinematic domain of the measurement is defined by $20^\circ < \theta^{\ell_{1,2}} < 150^\circ$, $P_T^{\ell_1} > 10$~GeV,  $P_T^{\ell_2} > 5$~GeV, $Q^2 < 1$~GeV$^2$, $y < 0.82$ and  $D^{\ell_1,\ell_2}_{\eta-\phi}>0.5$, where $D^{\ell_1,\ell_2}_{\eta-\phi}$ is the distance in the $\eta-\phi$ plane between the two
leptons. The effect of the $D^{\ell_1,\ell_2}_{\eta-\phi}$ requirement is small ($<1\%$).
The data samples at $\sqrt{s} = 301$~GeV and $319$~GeV are combined.
Assuming a linear dependence of the cross section on the proton beam energy, as predicted by the SM, the resulting cross section corresponds to an effective $\sqrt{s} = 318$~GeV.
The effect of final-state radiation on the cross sections was found to be negligible.

The total numbers of observed  $(\gamma\gamma)_{e}$ and $(\gamma\gamma)_{\mu}$ events are in agreement with the SM expectations, as summarised in Table~\ref{tab:mlepyields}. 
In the $(\gamma\gamma)_{e}$ sample, the contamination from NC DIS and QEDC background events is  $2$\%. No significant background is present in the $(\gamma\gamma)_{\mu}$ sample.
The contribution from $\tau$ pair production is negligible in both the $(\gamma\gamma)_{e}$  and $(\gamma\gamma)_{\mu}$ subsamples.
All high mass and high $\sum{P_T}$ events previously discussed are discarded from this sample by the requirement $E-P_z < 45~{\rm GeV}$.

The total visible and differential cross sections for electron and muon pair production are evaluated bin by bin as the weighted mean of the values measured by the two collaborations. 
The same binning is used by both experiments. 
The signal acceptance is defined as the number of events reconstructed in a bin divided by the number of events generated in the same bin and is calculated using GRAPE MC events. 
For  \mbox{$ep \rightarrow e \: e^+e^- X$} events, the mean signal acceptances in the H1 and ZEUS experiments are $45$\% and $60$\%, respectively. In case of \mbox{$ep \rightarrow e \: \mu^+\mu^- X$} events, it is $60$\% for H1 and $30$\% for ZEUS.

The total visible $ep \rightarrow ee^+e^- X$ cross section is  $\sigma = 0.68 \pm 0.04 \pm 0.03$~pb, where the first uncertainty is statistical and the second systematic. 
The total visible \mbox{$ep \rightarrow e\mu^+\mu^- X$} cross section is \mbox{$\sigma = 0.63 \pm 0.05 \pm 0.06$~pb}.
The results are in agreement with the SM expectation, dominated by photon-photon collisions,  of $0.69 \pm 0.02$~pb calculated using the GRAPE generator.
Since the muon and electron cross sections are compatible, as expected, they are combined into a single measurement, leading to a measured lepton pair production cross section of \mbox{$\sigma = 0.66 \pm 0.03 \pm 0.03$~pb}.
This result is in agreement with the individual H1 and ZEUS measurements~\cite{mlep_H1, mlep_ZEUS}.

\begin{figure}[htbp] 
\begin{center}
\includegraphics[width=\textwidth]{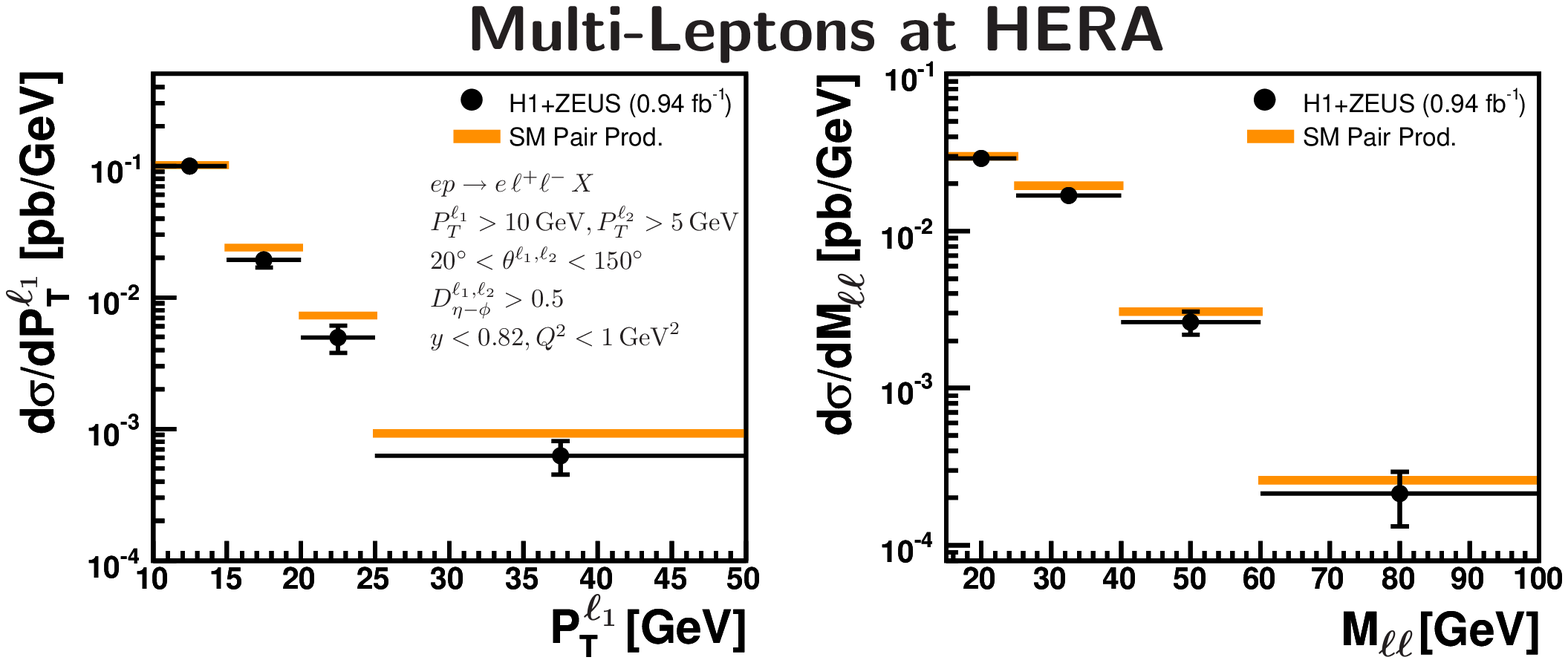}\put(-143,17){{(a)}}\put(-62,17){{(b)}}
\end{center}
  \caption{The cross section for lepton pair photoproduction in a restricted phase space as a function of the leading lepton transverse momentum $P_T^{\ell_1}$ (a) and the invariant mass of the lepton pair $M_{\ell\ell}$ (b).
The total error bar is shown,  representing the  statistical and systematic uncertainties added in quadrature, which is dominated by the statistical.
The bands represent the one standard deviation uncertainty in the SM prediction, dominated by the photon-photon process.   }
\label{fig:XSec_comb}
\end{figure}

Differential cross sections of lepton pair production as a function of the transverse momentum of the leading lepton $P_T^{\ell_1}$ and of the invariant mass of the lepton pair $M_{\ell\ell}$ are listed for each sample in Table~\ref{tab:xsection}  and shown in Fig.~\ref{fig:XSec_comb} for the combined electron and muon samples.
The measurements are in good agreement with the SM predictions.

\section{Conclusion}

The production of multi-lepton (electron or muon) events at high transverse momenta was studied using the full $e^\pm p$ data sample collected by the H1 and ZEUS experiments at HERA, corresponding to a total integrated luminosity of $0.94$~fb$^{-1}$.
The yields of di-lepton and tri-lepton events are in good agreement with the SM predictions.
Distributions of the invariant mass $M_{12}$ of the two highest $P_T$ leptons and of the scalar sum of the lepton transverse momenta $\sum P_T$ are in good overall agreement with the SM expectation. 

Events are observed in $ee$, $\mu\mu$, $e\mu$, $eee$ and $e\mu\mu$ topologies with invariant masses $M_{12}$ above $100$~GeV, where the SM expectation is low. 
Both experiments observe high mass and high $\sum P_T$ events in $e^+p$ collisions only, while, for comparable SM expectations, none are observed in $e^- p$ collisions.
Seven events have a \mbox{$\sum P_T > 100$~GeV}, whereas the corresponding SM expectation for $e^+p$ collisions is \mbox{$1.94 \pm 0.17$}. 

The total and differential cross sections for electron and muon pair photoproduction are measured in a restricted phase space dominated by photon-photon interactions.
The measured cross sections are in agreement with the SM predictions. 
%

\section*{Acknowledgements}

We are grateful to the HERA machine group whose outstanding
efforts have made these experiments possible.
We appreciate the contributions to the construction and maintenance of the H1 and ZEUS detectors of many people who are not listed as authors.
We thank our funding agencies for financial 
support, the DESY technical staff for continuous assistance and the 
DESY directorate for their support and for the hospitality they extended to the non-DESY members of the collaborations.


\clearpage

\begin{table}[]
\begin{center}
\begin{tabular}{ c c c c c }
\multicolumn{5}{c}{Multi-Leptons at HERA ($0.94$ fb$^{-1}$)}\\
\hline
Sample & Data & SM & Pair Production (GRAPE) & NC DIS + QEDC \\
\hline      
                                  
$ee$ & $873$ & $895 \pm 57$ & $724 \pm 41$  & $171 \pm 28$ \\ 
$\mu\mu$ & $298$   & $320 \pm 36$ & $320 \pm 36$ &  $< 0.5$ \\
$e\mu$ & $173$   & ~~$167 \pm 10$~~ & $152 \pm 9$~~ & ~$15 \pm 3$~ \\
$eee$ & $116$ & $119 \pm 7$~~ & ~$117 \pm 6$~~~ & $< 4$ \\    
$e\mu\mu$ & $140$ & $147 \pm 15$  & $147 \pm 15$  &  $< 0.5$    \\

\hline
$(\gamma\gamma)_{e}$ & $284$  & $293 \pm 18$ & $289 \pm 18$ & ~$4 \pm 1$\\ 
$(\gamma\gamma)_{\mu}$ & $235$ & $247 \pm 26$ & $247 \pm 26$ & $< 0.5$  \\  
\hline
\end{tabular}
\end{center}
\caption{Observed and predicted event yields for the different event topologies and for the $\gamma\gamma$ subsamples.
  The uncertainties on the predictions include model uncertainties and experimental systematic uncertainties added in quadrature. The limits on the background estimations
are quoted at $95$\% confidence level.}
\label{tab:mlepyields}
\end{table}

\begin{table}[]
\begin{center}
\begin{tabular}{ c c c c c }
\multicolumn{5}{c}{Multi-Leptons at HERA ($0.94$ fb$^{-1}$)}\\
\hline
\multicolumn{5}{c}{$M_{12}>$$100$~GeV}\\
\hline
Sample & Data & SM & Pair Production (GRAPE) & NC DIS + QEDC \\
\hline                                        
\multicolumn{5}{c}{$e^+p$ collisions ($0.56$ fb$^{-1}$)}\\
\hline                                        
$ee$  & $4$ & $1.68 \pm 0.18$ & $0.94 \pm 0.11$  & $0.74 \pm 0.12$ \\ 
$\mu\mu$  & $1$  & $0.32 \pm 0.08$ & $0.32 \pm 0.08$ &  $< 0.01$  \\ 
$e\mu$  & $1$	& $0.40 \pm 0.05$ & $0.39 \pm 0.05$ & $< 0.02$  \\ 
$eee$	& $4$ & $0.79 \pm 0.09$ & $0.79 \pm 0.09$ &  $< 0.03$    \\	
$e\mu\mu$ & $2$ & $0.16 \pm 0.04$ & $0.16 \pm 0.04$ &  $< 0.01$    \\
\hline                                        
\multicolumn{5}{c}{$e^-p$ collisions ($0.38$ fb$^{-1}$)}\\
\hline                                        
$ee$  & $0$ & $1.25 \pm 0.13$ & $0.71 \pm 0.11$ & $0.54 \pm 0.08$ \\ 
$\mu\mu$  & $0$  & $0.23 \pm 0.10$ & $0.23 \pm 0.10$ &  $< 0.01$  \\ 
$e\mu$  & $0$	& $0.26 \pm 0.03$ & $0.25 \pm 0.03$  & $< 0.02$  \\ 
$eee$	& $0$ & $0.49 \pm 0.07$ & $0.49 \pm 0.07$ &  $< 0.03$    \\	
$e\mu\mu$  & $0$ & $0.14 \pm 0.05$ & $0.14 \pm 0.05$ &  $< 0.01$    \\
\hline  				      
\multicolumn{5}{c}{All data ($0.94$ fb$^{-1}$)}\\
\hline
$ee$  & $4$ & $2.93 \pm 0.28$ & $1.65 \pm 0.16$  & $1.28 \pm 0.18$ \\ 
$\mu\mu$  & $1$  & $0.55 \pm 0.12$ & $0.55 \pm 0.12$ &  $< 0.01$  \\ 
$e\mu$  & $1$   & $0.65 \pm 0.07$ & $0.64 \pm 0.06$ & $< 0.02$  \\ 
$eee$   & $4$ & $1.27 \pm 0.12$ & $1.27 \pm 0.12$ &  $< 0.03$    \\    
$e\mu\mu$   & $2$ & $0.31 \pm 0.06$ & $0.31 \pm 0.06$ &  $< 0.01$    \\

\hline

\hline
\end{tabular}
\end{center}
\caption{Observed and predicted multi-lepton event yields for masses $M_{12} > 100$~GeV for the different event topologies, for all 
  data and divided into $e^+p$ and $e^-p$ collisions. 
  The uncertainties on the predictions include model uncertainties and experimental systematic uncertainties added in quadrature. The limits on the background estimations correspond to the selection of no event in the simulated topology and are quoted at $95$\% confidence level.}
\label{tab:mlepyieldsM100}
\end{table}

\begin{table}[]
\begin{center}
\begin{tabular}{ c c c c c }
\multicolumn{5}{c}{Multi-Leptons at HERA ($0.94$ fb$^{-1}$)}\\
\hline
\multicolumn{5}{c}{$\sum P_T>$$100$ GeV}\\
\hline
Data sample & Data & SM & Pair Production (GRAPE) & NC DIS + QEDC \\
\hline                                        
e$^{+}$p ($0.56$ fb$^{-1}$)  & $7$ & $1.94 \pm 0.17$ & $1.52 \pm 0.14$  & $0.42 \pm 0.07$ \\ 
e$^{-}$p ($0.38$ fb$^{-1}$)  & $0$ & $1.19 \pm 0.12$ & $0.90 \pm 0.10$  & $0.29 \pm 0.05$ \\ 
All      ($0.94$ fb$^{-1}$)  & $7$ & $3.13 \pm 0.26$ & $2.42 \pm 0.21$  & $0.71 \pm 0.10$ \\ 
\hline                                        
\end{tabular}
\end{center}
\caption{Observed and predicted multi-lepton event yields for $\sum P_T >$ $100$~GeV. Di-lepton and tri-lepton events are combined.
  The uncertainties on the predictions include model uncertainties and experimental systematic uncertainties added in quadrature.}
\label{tab:mlepyieldsEt100}
\end{table}

%
\begin{center}
 \renewcommand{\arraystretch}{1.15} 
 \begin{table}[]
\begin{center}
\begin{tabular}{ l l l l l l l l l l l l  }
\multicolumn{12}{c}{Multi-Leptons at HERA ($0.94$~fb$^{-1}$) }\\
\hline
\multicolumn{1}{c}{Variable}  &  \multicolumn{3}{c}{Measured} & \multicolumn{3}{c}{Measured}        &    \multicolumn{3}{c}{Measured}  & \multicolumn{2}{c}{Pair Production} \\
\multicolumn{1}{c}{ range}    &   \multicolumn{3}{c}{  ($e^+e^-$) }  &   \multicolumn{3}{c}{ ($\mu^+\mu^-$)}  &    \multicolumn{3}{c}{(average)}  &  \multicolumn{2}{c}{ (GRAPE)}     \\
\multicolumn{1}{c}{\mbox{[GeV]}} & \multicolumn{3}{c}{ [fb/GeV] }  &  \multicolumn{3}{c}{[fb/GeV] }  &   \multicolumn{3}{c}{[fb/GeV]}    &   \multicolumn{2}{c}{[fb/GeV]}   \\
\hline
\multicolumn{1}{c}{$P_T^{\ell_1}$} & \multicolumn{10}{c}{ $d\sigma/dP_T^{\ell_1}$}\\
\hline
$[10,15]$ & ~~$101.1$&$\!\!\!\!\!\pm \, 7.1$&$\!\!\!\!\!\pm \, 5.5$&~~~~$97.7$&$\!\!\!\!\!\pm \, 7.7$&$\!\!\!\!\!\pm \, 9.2$&~~~~$99.9$&$\!\!\!\!\!\pm \, 5.3$&$\!\!\!\!\!\pm \, 4.9$&~~$101.3$&$\!\!\!\!\!\pm \, 3.1$ \\
$[15,20]$ & ~~~~$22.4$&$\!\!\!\!\!\pm \, 3.1$&$\!\!\!\!\!\pm \, 1.3$&~~~~$15.9$&$\!\!\!\!\!\pm \, 3.2$&$\!\!\!\!\!\pm \, 1.7$&~~~~$19.4$&$\!\!\!\!\!\pm \, 2.3$&$\!\!\!\!\!\pm \, 1.0$&~~~~$23.9$&$\!\!\!\!\!\pm \, 0.7$ \\
$[20,25]$ & ~~~~~~$5.0$&$\!\!\!\!\!\pm \, 1.5$&$\!\!\!\!\!\pm \, 0.6$&~~~~~~$4.9$&$\!\!\!\!\!\pm \, 1.6$&$\!\!\!\!\!\pm \, 0.6$&~~~~~~$5.0$&$\!\!\!\!\!\pm \, 1.1$&$\!\!\!\!\!\pm \, 0.4$&~~~~~~$7.3$&$\!\!\!\!\!\pm \, 0.2$ \\ 
$[25,50]$ & ~~~~$0.56$&$\!\!\!\!\!\pm \, 0.22$&$\!\!\!\!\!\pm \, 0.05$&~~~~$0.75$&$\!\!\!\!\!\pm \, 0.29$&$\!\!\!\!\!\pm \, 0.09$&~~~~$0.63$&$\!\!\!\!\!\pm \, 0.18$&$\!\!\!\!\!\pm \, 0.04$&~~~~$0.93$&$\!\!\!\!\!\pm \, 0.03$ \\ %
\hline
\multicolumn{1}{c}{$M_{\ell\ell}$}  & \multicolumn{10}{c}{ $d\sigma/dM_{\ell\ell}$}\\
\hline
$[15,25]$   & ~~~~$27.3$&$\!\!\!\!\!\pm \, 2.8$&$\!\!\!\!\!\pm \, 1.5$&~~~~$31.9$&$\!\!\!\!\!\pm \, 2.9$&$\!\!\!\!\!\pm \, 3.0$&~~~~$29.0$&$\!\!\!\!\!\pm \, 2.1$&$\!\!\!\!\!\pm  \,1.5$&~~~~$30.0$&$\!\!\!\!\!\pm  \,0.9$ \\
$[25,40]$   & ~~~~$18.4$&$\!\!\!\!\!\pm  \,1.6$&$\!\!\!\!\!\pm \, 1.1$&~~~~$14.9$&$\!\!\!\!\!\pm \, 1.8$&$\!\!\!\!\!\pm  \,1.4$&~~~~$16.9$&$\!\!\!\!\!\pm \, 1.2$&$\!\!\!\!\!\pm  \,0.9$&~~~~$19.5$&$\!\!\!\!\!\pm \, 0.6$ \\
$[40,60]$   & ~~~~~~$3.4$&$\!\!\!\!\!\pm \, 0.6$&$\!\!\!\!\!\pm \, 0.2$&~~~~~~$2.0$&$\!\!\!\!\!\pm \, 0.5$&$\!\!\!\!\!\pm \, 0.2$&~~~~~~$2.6$&$\!\!\!\!\!\pm \, 0.4$&$\!\!\!\!\!\pm \, 0.2$&~~~~~~$3.1$&$\!\!\!\!\!\pm \, 0.1$ \\ 
$[60,100]$  & ~~~~$0.17$&$\!\!\!\!\!\pm \, 0.09$&$\!\!\!\!\!\pm \, 0.03$&~~~~$0.32$&$\!\!\!\!\!\pm \, 0.15$&$\!\!\!\!\!\pm \, 0.04$&~~~~$0.21$&$\!\!\!\!\!\pm \, 0.08$&$\!\!\!\!\!\pm \, 0.02$&~~~~$0.26$&$\!\!\!\!\!\pm  \,0.01$ \\ 
\hline                                        
\end{tabular}
\end{center}
\caption{Differential photoproduction cross sections $d\sigma/dP_T^{\ell_1}$ and $d\sigma/dM_{\ell\ell}$ averaged for each quoted interval for the process $ep \rightarrow e \ell^+ \ell^- X$ in a restricted phase space (see text for details). 
Cross sections are measured for $e^+e^-$ or $\mu^+\mu^-$ pairs.  The average is also shown.
The first uncertainty is statistical and the second is systematic.
Theoretical predictions, calculated with GRAPE, dominated by the photon-photon process, are shown in the last column.}
\label{tab:xsection}
\end{table}
\end{center}

\vfill
\newpage

\vfill
\newpage

\end{document}